\documentclass[10pt,aps, twocolumn,superscriptaddress,footinbib]{revtex4}

\usepackage[T1]{fontenc}
\usepackage[latin9]{inputenc}
\usepackage[letterpaper]{geometry}
\usepackage{float}
\usepackage{ae} 
\usepackage{color}
\usepackage[usenames,dvipsnames]{xcolor}

\usepackage{morefloats}
\usepackage{graphicx,subfigure,epsfig,epstopdf}
\usepackage{dcolumn}

\usepackage{amsmath,amssymb,mathtools}
\usepackage{fmtcount}
\usepackage{upgreek}

\input{epsf}
\input{epsfx}

\def\bra#1{\mathinner{\langle{#1}|}}
\def\ket#1{\mathinner{|{#1}\rangle}}
\def\braket#1{\mathinner{\langle{#1}\rangle}}
\def\Bra#1{\left<#1\right|}
\def\Ket#1{\left|#1\right>}

\begin{document}

\title{High-dimensional quantum key distribution using dispersive optics}

\author{Jacob Mower}
\affiliation{Research Laboratory of Electronics, Massachusetts Institute of Technology, Cambridge, MA 02139 USA}
\affiliation{Department of Electrical Engineering, Columbia University, New York, NY 10027 USA}
\author{Zheshen Zhang}
\affiliation{Research Laboratory of Electronics, Massachusetts Institute of Technology, Cambridge, MA 02139 USA}
\author{Pierre Desjardins}
\affiliation{Department of Applied Physics and Applied Mathematics, Columbia University, New York, NY 10027 USA}
\author{Catherine Lee}
\affiliation{Research Laboratory of Electronics, Massachusetts Institute of Technology, Cambridge, MA 02139 USA}
\affiliation{Department of Physics, Columbia University, New York, NY 10027 USA}
\author{Jeffrey H. Shapiro}
\affiliation{Research Laboratory of Electronics, Massachusetts Institute of Technology, Cambridge, MA 02139 USA}
\author{Dirk Englund}
\affiliation{Research Laboratory of Electronics, Massachusetts Institute of Technology, Cambridge, MA 02139 USA}
\affiliation{Department of Electrical Engineering, Columbia University, New York, NY 10027 USA}
\affiliation{Department of Applied Physics and Applied Mathematics, Columbia University, New York, NY 10027 USA}

\begin{abstract}

We propose a high-dimensional quantum key distribution protocol that employs temporal correlations of entangled photons. The security of the protocol relies on measurements by Alice and Bob in one of two conjugate bases, implemented using dispersive optics. We show that this dispersion-based approach is secure against collective attacks. The protocol is additionally compatible with standard fiber telecommunications channels and wavelength division multiplexers. We describe several physical implementations to enhance the transmission rate and describe a heralded qudit source that is easy to implement and enables secret-key generation at $>4$ bits per character of distilled key across over $200$ km of fiber.

\end{abstract}

\maketitle

\section{Introduction}

Quantum key distribution (QKD) \cite{BB84,1991.PRL.Ekert} enables two parties, Alice and Bob, to establish a private, shared cryptographic key. However, hardware constraints such as the optical state generation and photon-counting rates limit the rate of generating the key. By measuring photons in a high-dimensional Hilbert space, Alice and Bob may increase the shared information generated for each detected photon (or photon pair, in the case of entangled-photon QKD), thereby enabling greater key generation rates compared to measurements in low-dimensional Hilbert spaces. Furthermore, using high-dimensional correlations may provide greater robustness to noise \cite{2002.PRL.Cerf-Gisin.d-levelQKD}. Numerous degrees of freedom of photons have been investigated, including position-momentum \cite{PhysRevLett.100.110504}, time \cite{PhysRevA.73.031801,PhysRevA.61.062308,PhysRevLett.98.060503,PhysRevLett.84.4737}, energy-time \cite{Qi:06,DWDM_arxiv}, and orbital angular momentum (OAM) \cite{PhysRevLett.89.240401, Mair:2001fk}, but to our knowledge, no security proofs for these protocols have been published against collective or coherent attacks. 

Here, we introduce a high-dimensional QKD protocol that employs timing information of photons --- analogous to pulse position modulation (PPM) ---  to maximize the secret-key capacity under technical constraints. We focus the discussion on a scheme employing entangled photon pairs generated by Alice at random times by spontaneous parametric down-conversion (SPDC) and sent to Bob over a quantum channel, and we discuss variations of the scheme that  employ  single-photon sources or weak classical light. For the entangled-photon protocol, we show security against collective attacks through measurements by Alice and Bob in two conjugate bases, which are implemented using single-photon detectors and simple dispersive optical elements. This protocol, which we term dispersive optics QKD (DO-QKD), benefits from the robustness of temporal correlations in single-mode fiber and free space. We estimate that practical implementations could reach a secret-key capacity of $>4$ bits per character of distilled key (bpc) with transmission across over $200$ km in fiber. 

\section{The Protocol}

\begin{figure}
\centering\includegraphics[scale=0.51]{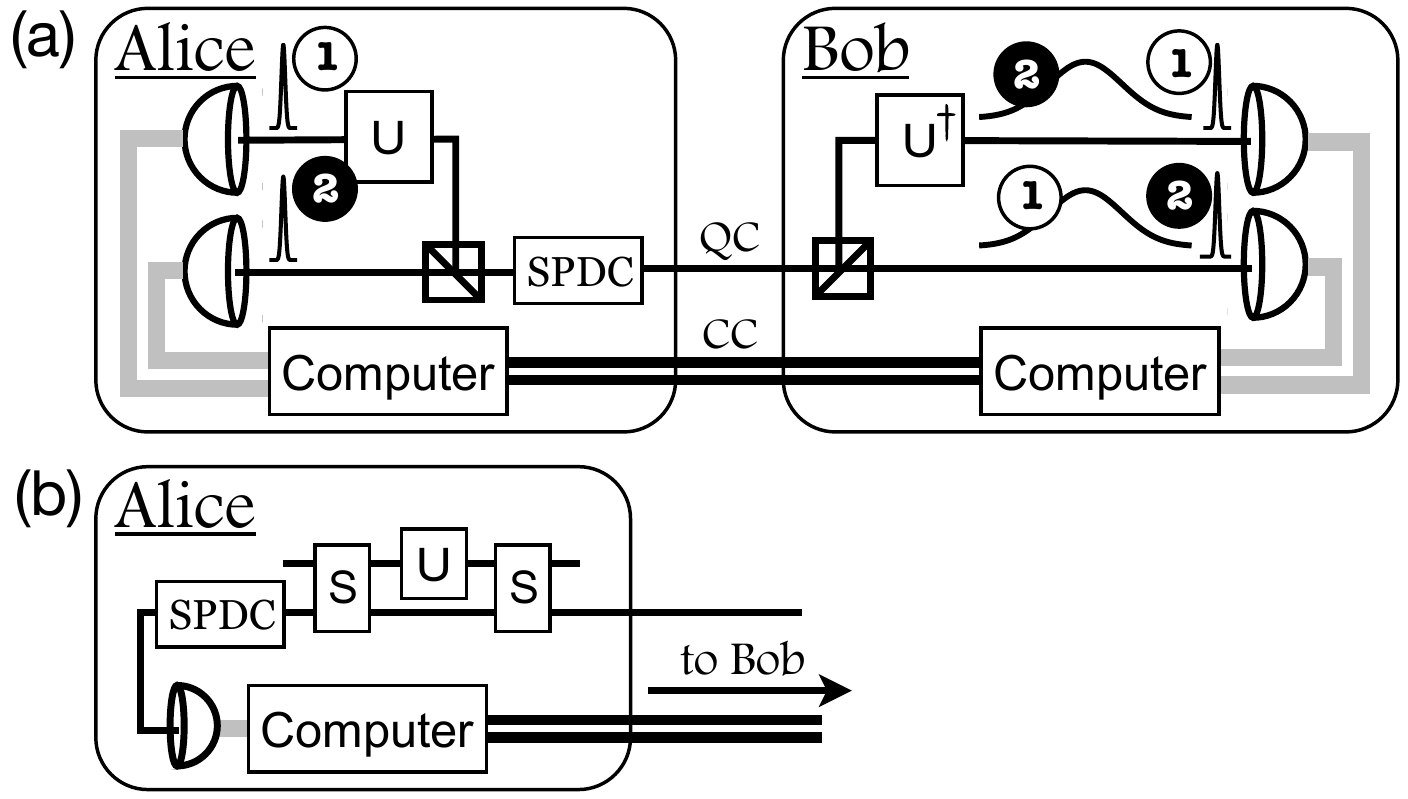}
\centering\caption{\small{(a) Alice and Bob measure in either the time or dispersed-time basis. In case (1), Alice measures in the dispersed-time basis and projects Bob's photon onto a dispersed state. In case (2), she projects Bob's photon onto an undispersed state. Only measurements in the same basis are correlated. QC represents quantum communication and CC represents classical communication. (b) An equivalent prepare-and-measure scheme in which Alice uses the arrival time of one photon as a sync time sent to Bob. She selects with a switch (S) whether or not to apply dispersion on her other photon and modulates the center time of the distribution that she sends to Bob using a Gaussian-distributed random number generator.}}
\label{fig1}
\end{figure}

The principal steps for the protocol are state preparation and transmission, state detection, and classical information post-processing. We present a schematic for these steps in Fig. \ref{fig1}(a).

\emph{1.) State preparation and transmission:} Alice generates a biphoton state via spontaneous parametric down-conversion (SPDC). For a weak, continuous-wave pump at frequency $\omega_p$ and operation at frequency degeneracy, the down-converted state (cf. Ref. \cite{PhysRevLett.98.060503}) can be approximated by 
\begin{equation}\vspace{-6pt}\label{eq:psiAB}
\ket{\Psi_{AB}}\!=\!\!\int\!\!\!\int\!\psi(t_{A},t_{B})e^{i \frac{\omega_p}{2}(t_A+t_B)}\ket{t_A t_B}dt_A dt_B,
\end{equation}
where
\begin{equation}
\psi(t_A,t_B)\!\propto\!e^{-(t_A-t_B)^2/4\sigma^2_{\text{cor}}}e^{-(t_A+t_B)^2/16\sigma^2_{\text{coh}}},\nonumber\\
\end{equation}
$\ket{t_A,t_B}=\hat{a}_A^{\dagger}(t_A)\hat{a}_B^{\dagger}(t_B)\ket{0}$, and $\hat{a}_{A,B}^\dagger(t_j)$ denote the creation operators at time $t_j$ for Alice and Bob, respectively. The superposition of temporal states in Eq. \ref{eq:psiAB} occurs over the coherence time of the pump field, $\sigma_{\textnormal{coh}}$. The correlation time between photons, $\sigma_{\textnormal{cor}}$, is determined by the phase matching bandwidth of the SPDC source. $\sigma_{\textnormal{coh}}$ can be longer than a $\upmu$s for a diode laser, and $\sigma_{\textnormal{cor}}$ is on the order of hundreds of fs to several ps for typical SPDC sources \cite{Zhong:09}. The resulting number of information eigenstates or alphabet `characters' \cite{PhysRevLett.92.127903} given by the Schmidt number, approximately $d\equiv \sigma_{\textnormal{coh}}/\sigma_{\textnormal{cor}}$, can therefore be quite large \cite{PhysRevA.73.031801}. 


In this investigation, we assume that Alice transmits the state to Bob over a standard telecom-band optical fiber, and $\omega_p/2 = 2\pi c/(1560 \textnormal{ nm})$.

\emph{2.) State detection:} Similar to the parties in the BB84 protocol, Alice and Bob choose randomly to measure their photons' arrival times in one of two bases: the basis of direct arrival-time measurements or a conjugate basis. These conjugate basis measurements can be achieved using a transformation $\hat{U}$ that transforms an eigenstate of the `direct measurement basis' into a superposition of all such eigenstates. We find that such a transformation can be implemented easily using group velocity dispersion (GVD), or `second-order dispersion.' An element with GVD imparts on each frequency state a phase $\phi \propto\beta_2 \omega^2$. $\beta_2=\partial^2/\partial\omega^2\vert_{\omega_0}(n_{\textnormal{eff }}\! \omega/c)$, where $n_{\textnormal{eff}}$ is the effective index of the mode, $\omega$ is the detuning from the mode's center frequency $\omega_0$, and $c$ is the speed of light in vacuum \cite{Engin}. Physically, $\beta_2$ is proportional to the linear change in the group velocity as a function of frequency. The SOD operator, $\hat{U}$, is unitary and its frequency domain representation is diagonal. 

Classically, a transform-limited pulse spreads out in time in a dispersive medium because its frequency components move out of phase. However, Ref. \cite{PhysRevA.45.3126} showed that if the entangled photons pass through dispersive media, in the limit of large coherence time $\sigma_{\textnormal{coh}}$, the correlation time $\sigma_{\textnormal{cor}}$ becomes
\begin{equation}
\sigma'^2_{\textnormal{cor}} \approx \frac{1}{\sigma_{\textnormal{cor}}^2}\left[\sigma_{\textnormal{cor}}^4+(\beta_{2A} L_A+\beta_{2B} L_B)^2\right], 
\end{equation}
where $\beta_{2A}$ ($\beta_{2B}$) is the GVD introduced by Alice (Bob) over length $L_A$ ($L_B$). Now, suppose that $L_A=L_B=L$ and $\beta_{\textnormal{tot}}=\beta_{2A}+\beta_{2B}$. As $\beta_{\textnormal{tot}}$ increases, the temporal correlation between Alice's and Bob's photons decreases. However, $\sigma'_{\textnormal{cor}}=\sigma_{\textnormal{cor}}$ if $\beta_{2A}=-\beta_{2B}=\beta_2$, which requires that, to within a global phase, $\hat{U}_A=\int e^{-i\frac{1}{2}\beta_{2} L \omega^2}\ket{\omega}_A\bra{\omega}d\omega$ and $\hat{U}_B^\dagger=\int e^{-i\frac{1}{2}\beta_{2} L \omega^2}\ket{\omega}_B\bra{\omega}d\omega$, where $\omega$ is the frequency detuning from the center frequency of the biphoton pulse, and $|\omega\rangle_X$ denotes a single photon at frequency $\omega_p/2 + \omega$ in the setup of party $X$. If Alice applies normal dispersion, $\hat{U}_A$, on her photon, Bob can apply anomalous dispersion of equal magnitude on his photon to recover the temporal correlation between their photons. Thus, Alice's and Bob's measurements in the dispersed basis are also correlated, as required. In order for the measurement bases to be conjugate, $\beta_2 L\gg\sigma_{\textnormal{coh}}\sigma_{\textnormal{cor}}$. To generate this normal and anomalous dispersion, a variety of technologies can be used, including commercial fiber Bragg gratings, silicon photonic crystal waveguides \cite{Assefa:07}, or optical cavities \cite{806867}. 

\emph{3.) Classical information post-processing}: Alice and Bob build their distilled key from correlated timing events acquired in the same basis~\cite{PhysRevLett.98.060503}. They therefore communicate their basis measurements and keep only the results where they registered clicks in the same basis. They additionally only consider frames during which both obtained one detection event. Using the security procedure detailed below, they determine their information advantage over Eve. If this is greater than zero, they then apply error correction and privacy amplification \cite{PhysRevLett.77.2818} on their data set to establish some amount of secret key. 

\section{Security}

To verify the security of the DO-QKD protocol against collective attacks, we calculate the secret-key capacity \cite{Devetak08012005,PhysRevLett.97.190503} in terms of bpc, as
\begin{equation}
\Delta I = \beta I(A;B) - \chi(A;E),
\end{equation}
where $\beta$ is the reconciliation efficiency, $I(A;B)$ is the mutual information between Alice and Bob, and $\chi(A;E)$ is Eve's Holevo information about Alice's transmission \cite{holevo73}. 
Since the low-flux limit of the SPDC output (given in Eq. \ref{eq:psiAB}) is Gaussian
and Gaussian attacks are optimal collective attacks for a measured covariance matrix \cite{PhysRevLett.97.190502, PhysRevLett.97.190503}, we can calculate the secret-key capacity using a covariance matrix approach to establish an upper bound on Eve's information given collective attacks \cite{PhysRevA.59.1820, 0953-4075-37-2-L02, RevModPhys.84.621}. 

\subsection{The noiseless covariance matrix, $\Gamma$}

The covariance matrix $\Gamma$ can be formulated using the measurement operators $\hat{T}_A$ ($\hat{T}_B$) and $\hat{D}_A$ ($\hat{D}_B$) corresponding to measurements by Alice (Bob) in the arrival time and dispersed arrival time bases, respectively. 
We first define these operators using several simplifying assumptions \cite{PhysRevA.56.3425}: 
\begin{itemize}
 	\item Alice and Bob's photons in $\Ket{\Psi_{AB}}$ have a negligible zero-momentum component.
	\item Each photon in this state propagates in a single direction directly preceding detection.
	\item The timing resolution of Alice's and Bob's detectors is insufficient to observe the effect of photon energy delocalization from the photon position distribution \cite{MandelandWolf}.
\end{itemize}
We can then approximate the arrival time operators as
\begin{equation}
\hat{T}_{j}=\int t_j\Ket{t_j}\Bra{t_j}dt_j,
\end{equation}
where $j\in\{A,B\}$. The dispersed arrival-time measurement operator $\hat{D}_j$ is related to $\hat{T}_j$ by a similarity transformation according to the dispersion operator $\hat{U}$ and a normalization giving units of frequency:
\begin{eqnarray}
\hat{D}_j&=&\frac{1}{\beta_{2j}L}\hat{U}_j^{\dagger}\hat{T}_j\hat{U}_j
\label{eqn:DU}\\
\hat{U}&=&\frac{1}{\sqrt{\pi|k|}}\int\int e^{-i(t_{1}-t_{2})^{2}/k}\left|t_{1}\right\rangle \left\langle t_{2}\right|dt_{2}dt_{1},\nonumber
\end{eqnarray}
where $k=2\beta_2 L$. Note that $[\hat{T}_j,\hat{D}_j]=i$. 
$\Gamma$ is a four-by-four matrix composed of four two-by-two submatrices denoted by
\begin{equation}
\Gamma=
\left( \begin{array}{cc}
\gamma_{AA} & \gamma_{AB} \\
\gamma_{BA} & \gamma_{BB} \\
\end{array} \right),
\end{equation}
where, for example, the submatrix $\gamma_{AB}$ describes the covariance between the measurements of Alice and Bob and is given by \cite{RevModPhys.84.621}
\begin{equation}
\gamma_{AB}= \frac{1}{2}
\left( \begin{array}{cc}
\braket{\{\hat{T_A},\hat{T_B}\}} & \braket{\{\hat{T_A},\hat{D_B}\}} \\
\braket{\{\hat{D_A},\hat{T_B}\}} & \braket{\{\hat{D_A},\hat{D_B}\}} \\
\end{array} \right),
\end{equation}
assuming the detection times in the arrival time and dispersed arrival time bases are centered around time zero. The noiseless covariance matrix --- i.e., the covariance matrix calculated in the absence of Eve's intrusion, channel effects and Alice's and Bob's setup imperfections --- is therefore given by
\begin{eqnarray}
\gamma_{AA}&=&
\left( \begin{array}{cc}
\frac{u+v}{16} & -\frac{u+v}{8k} \\
-\frac{u+v}{8k} & \frac{(u+v)(4k^2+uv)}{4k^2uv} \\
\end{array} \right) \nonumber\\
\gamma_{AB}&=&\gamma_{BA}^T=
\left( \begin{array}{cc}
\frac{u-v}{16} & \frac{u-v}{8k} \\
-\frac{u-v}{8k} & -\frac{(u-v)(4k^2+uv)}{4k^2uv} \\
\end{array} \right) \nonumber\\
\gamma_{BB}&=&
\left( \begin{array}{cc}
\frac{u+v}{16} & \frac{u+v}{8k} \\
\frac{u+v}{8k} & \frac{(u+v)(4k^2+uv)}{4k^2uv}  \nonumber
\end{array} \right), \nonumber
\end{eqnarray} 
where $u=16\sigma_{\textnormal{coh}}^2$ and $v=4\sigma_{\textnormal{cor}}^2$. In the limit of large dispersion where $k\rightarrow\infty$,
\begin{equation}
\Gamma\approx\left[\begin{array}{cccc}
\frac{u+v}{16} & 0 & \frac{u-v}{16} & 0\\
0 & \frac{u+v}{uv} & 0 & -\frac{u-v}{uv}\\
\frac{u-v}{16} & 0 & \frac{u+v}{16} & 0\\
0 & -\frac{u-v}{uv} & 0 & \frac{(u+v)}{uv}
\end{array}\right]
\end{equation}
which is equivalent to the covariance matrix calculated from arrival time and spectral measurement operators.

In the absence of noise, Alice and Bob perform photon arrival-time measurements with outcomes described by Gaussian-distributed random variables $T_{A}$ and $T_{B}$, respectively. The dispersed arrival-time elements, as they appear in the covariance matrix, have been multiplied by $1/\beta_{2}^{2}L^{2}$ due to the normalization in Eq. \ref{eqn:DU}. Before adding any timing noise due to Eve or the transmission channel, we first multiply these elements by $\beta_{2}^{2}L^{2}$ to convert the normalized variances with units of frequency back to temporal variances. Therefore, from this point on, we will assume that the Gaussian-distributed random variables $D_{A}$ and $D_{B}$ are given in units of time.
\subsection{Covariance matrix $\Gamma'$ used to calculate $\chi(A;E)$}

We consider the effect of an eavesdropper and channel noise, which result in excess noise $\epsilon$ and a decrease in correlations $\eta$. Alice and Bob know how much noise is added to their measurements by dark counts and jitter and assume that Eve cannot control these sources of noise. Suppose that the temporal measurements by Alice and Bob, in the presence of Eve yield values $T_{A}'$, $T_{B}'$, $D_{A}'$, and $D_{B}'$. The variances of these primed values are related to the unprimed values (no Eve) according to
\begin{equation}
\textnormal{COV}[T_{A}',T_{B}']=(1-\eta)\textnormal{COV}[T_{A},T_{B}]
\end{equation}
\begin{equation}
\textnormal{Var}[T_{A}']=\textnormal{Var}[T_{A}]
\end{equation}
\begin{equation}
\textnormal{Var}[T_{B}']=\left(1+\epsilon\right) \textnormal{Var}[T_{B}].
\end{equation}
The primed dispersed arrival time variables are related to the unprimed variables in an analogous way. Excess noise appears only in in Bob's measurements because Alice's photons do not leave her setup (cf. Fig. \ref{fig1}(a)).

\subsection{Covariance matrix $\Gamma''$ used to calculate $I(A;B)$}
While imperfections in Alice's and Bob's setups do not contribute to $\chi(A;E)$, they do lower $I(A;B)$. If we also include detector timing jitter and dark counts, Alice's and Bob's arrival-time measurements are described by 
\begin{eqnarray}
T_{A}''= 
\begin{dcases}
    T_{A}'+N_{A}^{J}, & \text{with probability } R_{\nu A}\\
    N_{A}^{dT}, & \text{with probability } R_{dA}
\end{dcases}\\
T_{B}''= 
\begin{dcases}
    T_{B}'+N_{B}^{J}, & \text{with probability } R_{\nu B}\\
    N_{B}^{dT}, & \text{with probability } R_{dB}
\end{dcases}
\end{eqnarray}
where $N_{A/B}^{J}$ is the noise due to detector jitter, $N_{A/B}^{dT}$ is the noise due to dark counts when measuring the temporal variance, and $R_{\nu A/B}$ ($R_{dA/B}$) is the probability of registering a photon (dark count) given a single click on Alice's/Bob's detector. 

We calculate $R_{\nu}$ and $R_{d}$ as follows, ignoring for now the possibility of generating more than one photon pair per frame. The probability that Alice/Bob detects a photon and not a dark count $p_{A/B}$ is the probability that Alice's source generates a photon pair, a photon from the pair arrives at Alice's/Bob's detector, is detected, and a dark count is not registered. Therefore
\begin{equation}
p_{A/B}=p_\nu  (1-L_{A/B})(1-p_d).
\end{equation}
where $L_{A}$ ($L_{B}$) is the loss in Alice's (Bob's) detection system (including the channel for Bob), $p_d$ is the probability that Alice or Bob's detector registers a dark count in a frame, and $p_\nu $ is the probability of generating a pair in a given frame. The probability of either party registering a dark count given one detection event in a frame is
\begin{equation}
d_{A/B}=\left[p_\nu L_{A/B}+(1-p_\nu )\right]p_d.
\end{equation}
From these results, we have 
\begin{equation}
R_{\nu A/B}=\frac{p_{A/B}}{p_{A/B}+d_{A/B}}
\end{equation}
\begin{equation}
R_{dA/B}=\frac{d_{A/B}}{p_{A/B}+d_{A/B}}.
\end{equation}
From Eq. 9-13, it follows that
\begin{eqnarray}
\textnormal{COV}[T_{A}'',T_{B}''] \!&=&\! \textnormal{COV}[R_{\nu A}T_{A}',R_{\nu B}T_{B}']\\
 \!&=&\! R_{\nu A}R_{\nu B}(1-\eta)\textnormal{COV}[T_{A},T_{B}]\nonumber\\
 \text{Var}[T_A''] &=& R_{\nu A}(\text{Var}[T_{A}]+\text{Var}[N_A^J])\nonumber\\
                                              && +R_{dA}\text{Var}[N_A^{dT}]\\
 \text{Var}[T_B''] &=& R_{\nu B}(\text{Var}[T_B]+\text{Var}[N_B^J])\nonumber\\
                                              && +R_{dB}\text{Var}[N_B^{dT}].
\end{eqnarray}
Alice's and Bob's dispersed arrival-time variables obey corresponding relations. 
The variance of $N_{A/B}^{J}$ is $\sigma_J^{2}$. Since dark counts are uniformly distributed, if Alice and Bob measure the variance in the arrival-time and dispersed arrival-time bases to three standard deviations of $T_{A/B}'$ and $D_{A/B}'$, the variances of $N_{A/B}^{dT}$ and $N_{A/B}^{dD}$ are
\begin{equation}
\textnormal{Var}[N_{A/B}^{dD}]=\frac{1}{E}\int_{-E/2}^{E/2}x^{2}dx=\frac{1}{12}E^{2},
\end{equation}
\begin{equation}
\textnormal{Var}[N_{A/B}^{dT}]=\frac{1}{F}\int_{-F/2}^{F/2}x^{2}dx=\frac{1}{12}F^{2},
\end{equation}
where $E\equiv6\sqrt{\textnormal{Var}[D_{A}']}$ and $F\equiv6\sqrt{\textnormal{Var}[T_{A}']}$. These covariances can then be used to calculate the secret key capacity, as detailed in Appendix A.

\subsection{Noise parameters}

While it is convenient to use $\epsilon$ and $\eta$ to describe Eve's effect on the covariance matrix, it is easier experimentally to measure a different set of parameters. In particular, we consider $\xi$ and $\theta$ defined as
\begin{equation}
\text{Var}[T_A' - T_B']=(1+\xi)\text{Var}[T_A - T_B],
\end{equation}
and
\begin{equation}
\text{Var}[T_A' + T_B']=(1-\theta)\text{Var}[T_A + T_B].
\end{equation}
Both $\{\xi,\theta\}$ and $\{\epsilon,\eta\}$ can be used equivalently to bound Eve's information. These sets are related by
\begin{equation}
\epsilon = \frac{-2\eta(d^2-\frac{1}{4})+\xi}{d^2+\frac{1}{4}}.
\label{eq:epsetaxi}
\end{equation}
Therefore, if Alice and Bob measure only $\xi$ and consider all physical $\epsilon$ and $\eta$ according to Eq. \ref{eq:epsetaxi}, they can find a minimum secret-key capacity $\Delta I$. This procedure is practical if, for a given $\xi$, $\Delta I$ has only a weak dependence on $\eta$ and $\epsilon$. To illustrate why this is useful, consider $\epsilon=\eta=6.3\cdot10^{-5}$, $d=64$, $\xi=0.78$, and $\sigma_\text{cor}=30$ ps. $\xi$ corresponds to a change in the correlation time, $\sigma_\Delta = \sigma_\text{cor}(\sqrt{1+\xi}-1)=10$ ps. $\epsilon$ corresponds to an increase in the variance of Bob's measurements by $<0.1$ ps, which is more difficult to detect.

The range of physical values for $\epsilon$ and $\eta$ are given by a number of constraints: (i) Eve cannot increase the mutual information of Alice and Bob by interacting with only Bob's photons due to the data processing inequality; (ii) The symplectic eigenvalues of the covariance matrix (see Appendix \ref{sec:symp}) are greater than $\frac{1}{2}$; (iii) Eve can only degrade Alice and Bob's measured time correlation, i.e., $\text{Var}[T_A'-T_B']\ge \text{Var}[T_A-T_B]$. Under these constraints, we can then calculate an upper bound on Eve's information.

To relate noise measures for different $d$, we use $\sigma_\Delta$ in the following calculations.

\section{Key Rates}

While we have considered photons generated by SPDC, it may not be possible to experimentally determine the covariance matrix elements for such a state. In Eq.  \ref{eq:psiAB}, we assume that the biphoton amplitude is centered at times $\braket{t_A}=\braket{t_B}=0$. However, since the pairs are generated at random times under continuous-wave excitation, Alice and Bob do not know the center time of the biphoton envelope. Therefore, Alice and Bob could use a prepare-and-measure scheme, shown in Fig. \ref{fig1}(b), in which Alice directly measures the arrival time of her photon, sends this time to Bob as a synchronization pulse, and then modulates the center time of the distribution that she sends to Bob using a Gaussian-distributed random number generator.

In addition to enabling measurement of the covariance matrix, this technique allows Alice to increase the photon generation rate. Assuming Alice has high system detection efficiency, she can determine if multiple pairs are emitted in each frame and remove them with an amplitude modulator. This heralding and post-selection scheme \emph{allows Alice to send nearly one photon per frame with a low probability of sending multiple photons}. We find that the multi-pair probability can then be bounded below $0.01$, even when the expected pair generation rate per frame, $\mu_f\approx1$  (see derivation in Appendix \ref{sec:heralding}). This ability for efficient post-selection points to an important advantage of using high-dimensional encoding: for high $d$, the purity of the single photon source after post-selection increases for a given pump power. We will use this post-selection scheme now to analyze the DO-QKD protocol when it uses on the order of one photon per frame.
\begin{figure}[h!]
\centering\includegraphics[scale=0.38]{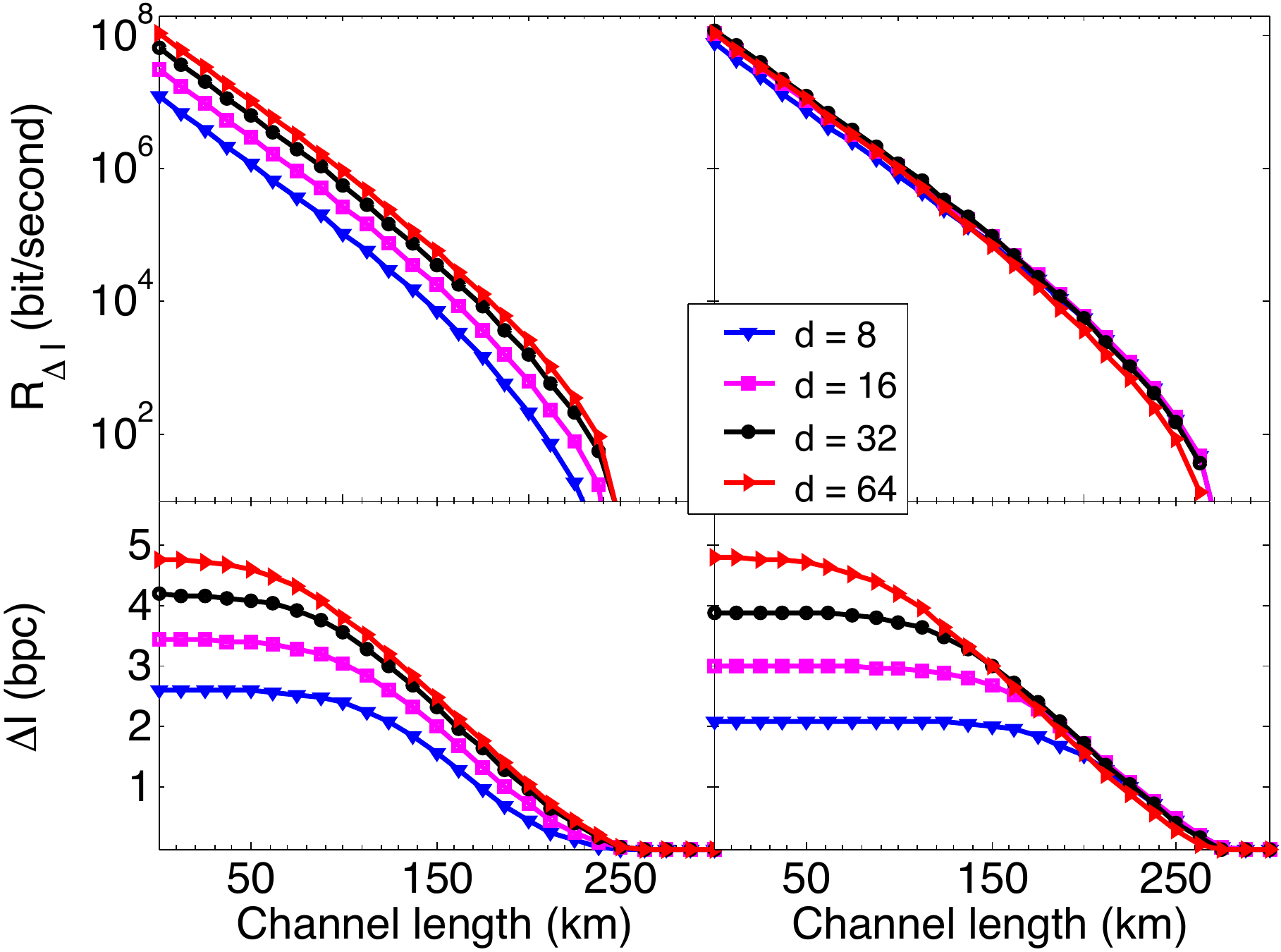}
\centering\caption{\small{(left) Secret-key capacity as a function of channel length, assuming equal photon pair generation rate $\gamma_\nu=p_\nu /6\sigma_{\textnormal{coh}}$ and $\sigma_{\textnormal{coh}}=64\cdot30$ ps for all $d$; propagation loss $\alpha=0.2$ dB/km; detector timing jitter $\sigma_J=20$ ps; Alice and Bob system detection efficiencies $93\%$; 
$\sigma_\Delta=10$ ps; dark count rate, $r_D=1000$ s$^{-1}$; $\beta=0.9$; expected number of pairs per frame assuming heralding, $p_\nu =\{0.607,0.411,0.231,0.119\}$ for $d=\{64, 32, 16, 8\}$, respectively (see Appendix \ref{sec:heralding}). (right) The secret-key capacity given the same parameters, but with $\gamma_\nu=p_\nu /(6d\cdot\sigma_{\textnormal{cor}})$ and $\sigma_{\textnormal{cor}}=30$ ps for all $d$. Upper plots show bps. Lower plots show bpc, i.e., bits per frame in which Alice and Bob measure in the same basis and register only one detection event.}}
\label{results}
\end{figure}

The secret-key rate, $R_{\Delta I}$, with units of secure bits per second (bps) is given by $R_{\Delta I}=\Delta I\cdot P_{C}\cdot \gamma_\nu$, where $\gamma_\nu=p_\nu /6\sigma_\text{coh}$, $P_{C}$ is the probability that both Alice and Bob register one click in a given frame, and $\Delta I$ is the secret-key capacity from Eq. 3. The factor of $6$ serves to separate the center time of photons in neighboring frames by more than $6$ standard deviations.

%
Even for collective attacks, Alice and Bob can share a large amount of information per second using the DO-QKD protocol. Pair generation rates in excess of $10^9$ s$^{-1}$ are possible using moderate pump powers \cite{Zhong:09}, enabling secure communication rates $> 100$ Mb/s. In Fig. \ref{results}, we plot $\Delta I$ and $R_{\Delta I}$ as a function of fiber channel length. With $\sigma_\Delta = 10$ ps, over $\sim200$ km transmission length can be achieved. 

\begin{figure}[h!]
\centering\includegraphics[scale=0.40]{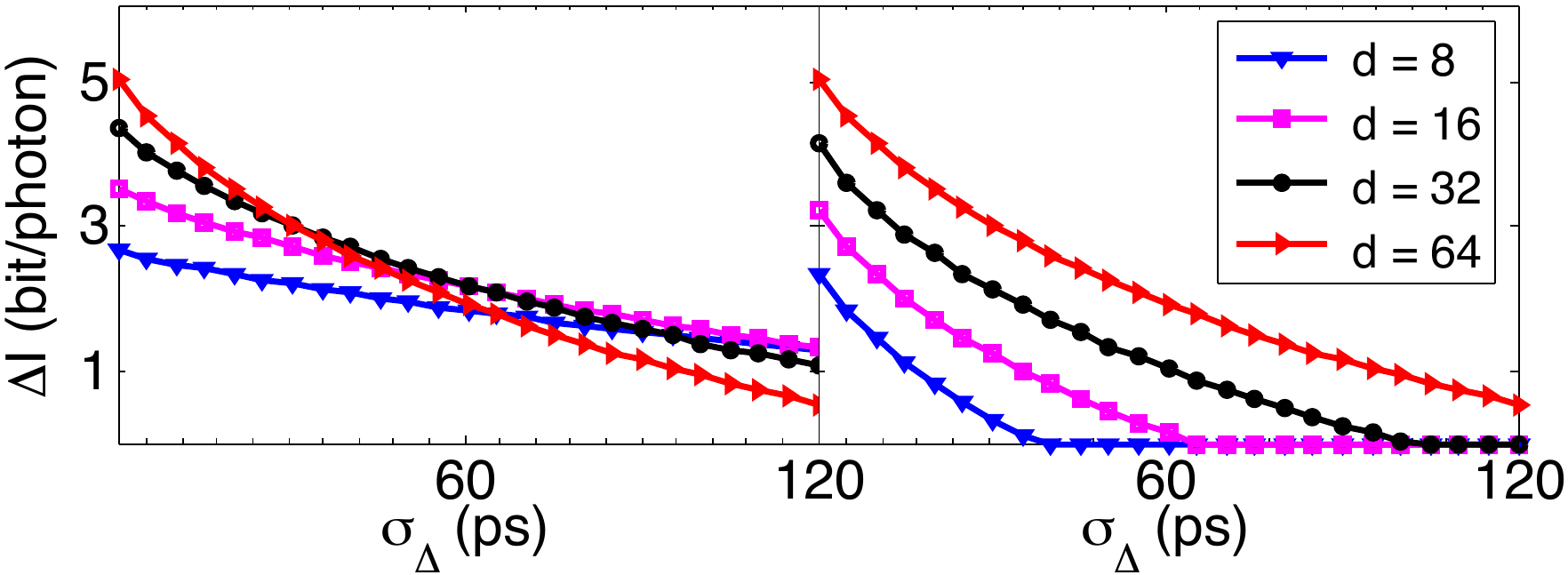}
\centering\caption{\small{$\Delta I$ as a function of offset parameter $\sigma_\Delta$ assuming the parameters given in the caption of Fig. \ref{results}. (left) Photon pair generation rate $\gamma_\nu=p_\nu /6\sigma_{\textnormal{coh}}$ and $\sigma_{\textnormal{coh}}=64\cdot30$ ps for all $d$. (right) $\gamma_\nu=p_\nu /(6d\cdot\sigma_{\textnormal{cor}})$ and $\sigma_{\textnormal{cor}}=30$ ps for all $d$.}} 
\label{options}
\end{figure}

Fig. \ref{options} plots the dependence of $\Delta I$ on $\sigma_\Delta$ for different $d$, assuming the parameters given in the Fig. \ref{results} caption. Even if Alice and Bob measure $\sigma_\Delta$ to be on the order of tens of picoseconds for $ \sigma_\text{cor}=30$ ps, they can extract a positive $\Delta I$. Since we assume detector timing jitter of $20$ ps, $\sigma_\Delta$ of this order is not difficult to measure. 

\section{Discussion}



We have so far considered a passive selection of basis measurements by Bob, using a beamsplitter followed by two single-photon detectors, as shown in Fig. 1(a). This number of detectors could be reduced to only one if the beam splitter were replaced by an active switch, as shown in Fig. 1(b) 
Additional detectors would be required to register instances in which multiple photons arrived at the detector within its reset time. In particular, detecting instances in which $m$ photon pairs are generated requires at least $m$ detectors. But this probability, which is given by the g-fold degenerate Bose-Einstein distribution (see Appendix B), rapidly diminishes with $m$. 
Thus, even for $\mu\sim 1$, only a few extra detectors would be required.  The DO-QKD protocol can therefore be very resource efficient, whereas other protocols that employ high-dimensional states in frequency \cite{Qi:06,DWDM_arxiv} or OAM modes \cite{PhysRevLett.89.240401, Mair:2001fk} require a number of detectors that grows linearly with the dimensionality $d$. 

It is interesting to consider why the DO-QKD protocol can reach a much longer transmission length than continuous variable QKD (CV-QKD) protocols, which are so far limited to $80$ km \cite{80km_cvqkd}. We believe that a primary reason for the shorter transmission length in CV-QKD is that $\epsilon$ and $\eta$ increase with photon loss/channel length. By contrast, in DO-QKD, $\epsilon$ and $\eta$ are constant with photon loss/channel length (we believe that this would be the case for any single photon QKD protocol employing continuous variables). Ultimately, the transmission length in DO-QKD is limited not by an increase in the Holevo information between Alice and Eve, but rather by a decrease in the mutual information between Alice and Bob; $\Delta I$ decreases to zero with increasing channel length as the probability of registering a dark count approaches the probability of Bob detecting a photon from Alice. 


Transmission $> 200$ km is realistic because the DO-QKD protocol is ideally suited for fiber telecommunications networks and photonic integrated circuits (PIC), which are designed for high-bandwidth temporal encoding. In Ref. \cite{PhysRevA.66.042303}, a protocol was proposed using PICs for the efficient generation of single photons on-demand \cite{PhysRevA.84.052326}, and PICs can be used to implement dispersive elements, as noted in Sec. II. 

To further reduce the complexity of Alice's setup, she could use an attenuated classical source instead of an entangled photon source. As in pulse position modulation (PPM), Alice would simply modulate the optical field to generate random code-words. Two switches could then be used for randomly transforming the PPM signals into one of the temporal bases before sending them onward to Bob (note that the loss introduced by these switches can be compensated by increasing the light intensity). Note that Alice no longer requires a photon detector in this implementation: she only needs a modulated light source and dispersive optics. This attenuated-light implementation could be paired with a decoy state protocol  \cite{PhysRevLett.94.230504} to protect against photon number splitting attacks. 

For channel bandwidths of $\sim 1/\sigma_{\textnormal{cor}}$, DO-QKD can operate at the Heisenberg measurement limit, while for larger bandwidths, dense wavelength division multiplexing (DWDM) could allow operation on hundreds of independent wavelength channels. 

\section{Conclusion}

We have introduced a scheme to perform high-dimensional QKD to maximize the key generation rate given experimental limitations such as detector reset time or finite brightness of entangled photon sources. Because the protocol relies on temporal correlations, it is ideally suited for fiber communication systems. It is also resource efficient and can be implemented with as few as two detectors using entangled photon pairs. If Alice used PPM and dispersion, only Bob would require a single photon detector. The DO-QKD protocol allows the generation of $>4$ bpc with communication rates exceeding $100$ Mb/s and communication distances over $200$ km. We show that this protocol is secure against collective attacks. Moreover, single photon heralding with high dimensionality $d$ can be used to increase the communication rate. Future work will focus on increasing the rate at which Alice and Bob can perform the necessary post-processing for secure communication, as this remains a rate-limiting step for single-photon QKD protocols.

This work was supported by the DARPA Information in a Photon program, through grant W911NF-10-1-0416 from the Army Research Office, by the Sloan Research Fellowship, and the Columbia Optics and Quantum Electronics IGERT under NSF grant DGE-1069420.

\appendix

\section{Symplectic decomposition of the covariance matrix}
\label{sec:symp}
The quantum state described in Eq. 1 is the post-selected, low-flux limit of the Gaussian state generated by spontaneous parametric down-conversion, and can be fully characterized by $\Gamma'$. Here, we consider the collective attack, in which Eve performs only individual interactions with the photons flowing to Bob, but makes a collective measurement on the joint state she derives from all such interactions.

The Holevo information under this attack can be found by \cite{PhysRevA.72.012332}
\begin{eqnarray}
\label{eqChiAE1}
\chi(A;E) \!\!&=&\!\! S(\rho_{E})-\frac{1}{2}(H_T +H_\Omega)\\
\label{eqChiAE2}
H_T \!\!&=&\!\! \int p(t_{A}=t)S(\rho_{E|t_{A}=t})dt\\
\label{eqChiAE3}
H_\Omega \!\!&=&\!\! \int p(\omega_{A}=\omega)S(\rho_{E|\omega_{A}=\omega})d\omega
\end{eqnarray}
where $S(\rho)$ is the von Neumann entropy of the quantum state $\rho$, $p(t_{A}=t)$ is the probability density for Alice to measure $t_{A}$ in the arrival-time basis, and $p(\omega_{A}=\omega)$ is the probability density for Alice to measure $\omega_{A}$ in the dispersed arrival-time basis. Since Alice, Bob, and Eve's overall quantum state $\rho_{ABE}$ is a pure state, $S(\rho_{E})=S(\rho_{AB})$. After Alice's measurement, the quantum state shared by Bob and Eve is pure. Therefore, $S(\rho_{E|t_{A}=t})=S(\rho_{B|t_{A}=t})$ and $S(\rho_{E|\omega_{A}=\omega})=S(\rho_{B|\omega_{A}=\omega})$. Furthermore, given the fact that all states are Gaussian, Bob and Eve's conditional quantum state is independent of Alice's measurement result. Thus, we can drop the integrals in Eqs. \ref{eqChiAE2} and \ref{eqChiAE3} since the integrand is a constant. Then Eq. \ref{eqChiAE1} becomes 
\begin{equation}
\chi(A;E)=S(\rho_{AB})-\frac{1}{2}\left[S(\rho_{B|t})+S(\rho_{B|\omega})\right].\label{eqChiAE_final}
\end{equation}
After Eve's interaction, Alice and Bob's covariance matrix becomes
\begin{equation}
\Gamma' = \left[\begin{array}{cc}
\gamma_{\text{AA}} & (1-\eta)\gamma_{\text{AB}}\\
(1-\eta)\gamma_{\text{BA}} & (1+\varepsilon)\gamma_{\text{BB}}
\end{array}\right].
\end{equation}
$\gamma_{\text{AA}}$ remains unchanged because Eve does not have access to Alice's photon. We define 
\begin{eqnarray}
I_1\!\!&=&\!\!\det[\gamma_{AA}]+\det[(1+\epsilon)\gamma_{BB}]+2\det[(1-\eta)\gamma_{AB}]\nonumber\\
I_2\!\!&=&\!\!\det[\Gamma']\\
d_{\pm}\!\!&=&\!\!\frac{1}{\sqrt{2}}\sqrt{I_{1}\pm\sqrt{I_{1}^{2}-4I_{2}}}\nonumber.
\end{eqnarray}
$S(\rho_{AB})$ is evaluated by $S(\rho_{AB})=f(d_{+})+f(d_{-})$, where 
\begin{equation}
f(x)=(x+\frac{1}{2})\log_{2}(x+\frac{1}{2})-(x-\frac{1}{2})\log_{2}(x-\frac{1}{2}).
\end{equation}

To calculate the conditional terms in Eq. \ref{eqChiAE_final}, we first need to derive Bob's conditional covariance matrices $\gamma_{B|t}$ and $\gamma_{B|\omega}$, which are given by 
\begin{subequations}
\begin{align}
 & \gamma_{B|t}=\gamma_{BB}-\gamma_{BA}\left(X_{t}\gamma_{AA}X_{t}\right)^{-1}\gamma_{AB}\\
 & \gamma_{B|\omega}=\gamma_{BB}-\gamma_{BA}\left(X_{\omega}\gamma_{AA}X_{\omega}\right)^{-1}\gamma_{AB},
\end{align}
\end{subequations} where $X_{t}=\left[\begin{array}{cc}
1 & 0\\
0 & 0
\end{array}\right]$, $X_{\omega}=\left[\begin{array}{cc}
0 & 0\\
0 & 1
\end{array}\right]$, 
and the matrix inverse is carried out by the Moore-Penrose pseudoinverse. The entropy of the conditional states reads 
\begin{subequations}
\begin{align}
 & S(\rho_{B|t})=f(\sqrt{\det[\gamma_{B|t}]})\\
 & S(\rho_{B|\omega})=f(\sqrt{\det[\gamma_{B|\omega}]}).
\end{align}
\end{subequations} 
We next calculate the mutual information between Alice and Bob. The classical mutual information $I(A;B)$ can be evaluated from $\Gamma''$ by 

\begin{equation}
I(A;B)=\frac{1}{4}\left[\log_{2}\left(\frac{1}{1-\mu_{T}^{2}}\right)+\log_{2}\left(\frac{1}{1-\mu_{D}^{2}}\right)\right],
\end{equation}
where $\mu_T$ and $\mu_D$ are the correlation coefficients for Alice and Bob's arrival-time and dispersed arrival-time measurements, given by
\begin{eqnarray}
\mu_{T}\!\!&=&\!\!\frac{\text{COV}[T_A'',T_B'']}{\sqrt{\text{Var}[T_A'']\text{Var}[T_B'']}}\label{eq:mut}\\
\mu_{D}\!\!&=&\!\!\frac{\text{COV}[D_A'',D_B'']}{\sqrt{\text{Var}[D_A'']\text{Var}[D_B'']}}\label{eq:mud}\nonumber
\end{eqnarray}
for $T_A''$ and $T_B''$ defined in the text.

\section{Optimizing the photon source and detectors}
\label{sec:heralding}

The security analysis in the text assumed that at most a single pair was generated by Alice in each frame. However, for an SPDC source, the probability of generating $m$ pairs over some time interval is given by g-fold degenerate Bose-Einstein statistics as \cite{Mandelphotstat}
\begin{equation}
p(\mu,m,g)=\frac{(\mu/g)^m/(g-1)}{[1+(\mu/g)^{m+g}]B(m+1,g-1)}
\end{equation}
where $\mu$ is the expected number of pairs generated during that interval, $g$ is the mode degeneracy and $B(x,y)$ is the Beta function. We take $g=d$ in our calculations. Thus to suppress the multi-pair emission probability, one must suppress $\mu$, reducing the probability of generating any photons at all. One can avoid this by employing a heralding scheme \cite{Yang:2011fk}, whereby the detection of Alice's photon heralds the existence of Bob's, and Alice blocks the channel such that no more than one signal photon leaves her setup per frame. Alice then randomizes the center time of the photon distribution sent to Bob according to a Gaussian distribution to generate the Gaussian measurement statistics assumed in our analysis. 

The relevant period over which to consider multi-photon events is the time bin instead of the protocol time frame, which can allow Alice to increase the pair generation rate while suppressing multi-pair generation.

\begin{figure}[t]
\centering\includegraphics[scale=0.78]{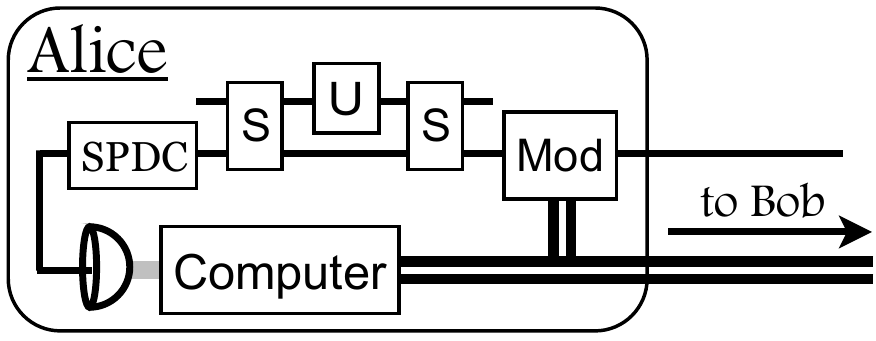}
\centering\caption{\small{A prepare-and-measure scheme in which Alice additionally modulates the output channel according to the heralding scheme outlined in the text using the element, Mod. Arrival time measurements from her idler photon are fed forward to Mod enabling Alice to suppress multiphoton emission during each frame.}}
\label{her_circ}
\end{figure}

The experimental setup is depicted in Fig. \ref{her_circ}. Alice prepares $\ket{\Psi_{AB}}$ and sends her photons directly to a single photon detector. Detection at this stage heralds the presence of Bob's photons. If she detects more than one photon per frame, she chooses one photon at random  to pass to Bob and modulates the SPDC output channel with a modulator (Mod) to remove all other photons. We assume 93\% detection efficiency \cite{sw-nam93} and 1 dB attenuation in the switch. The detection can occur at the required rate $> 100$ MHz assuming the use of detector arrays \cite{4277352}. Since the dark count rate is roughly six orders of magnitude lower than the pair generation rate, we can safely neglect them in these calculations.

\begin{figure}
\centering\includegraphics[scale=0.75]{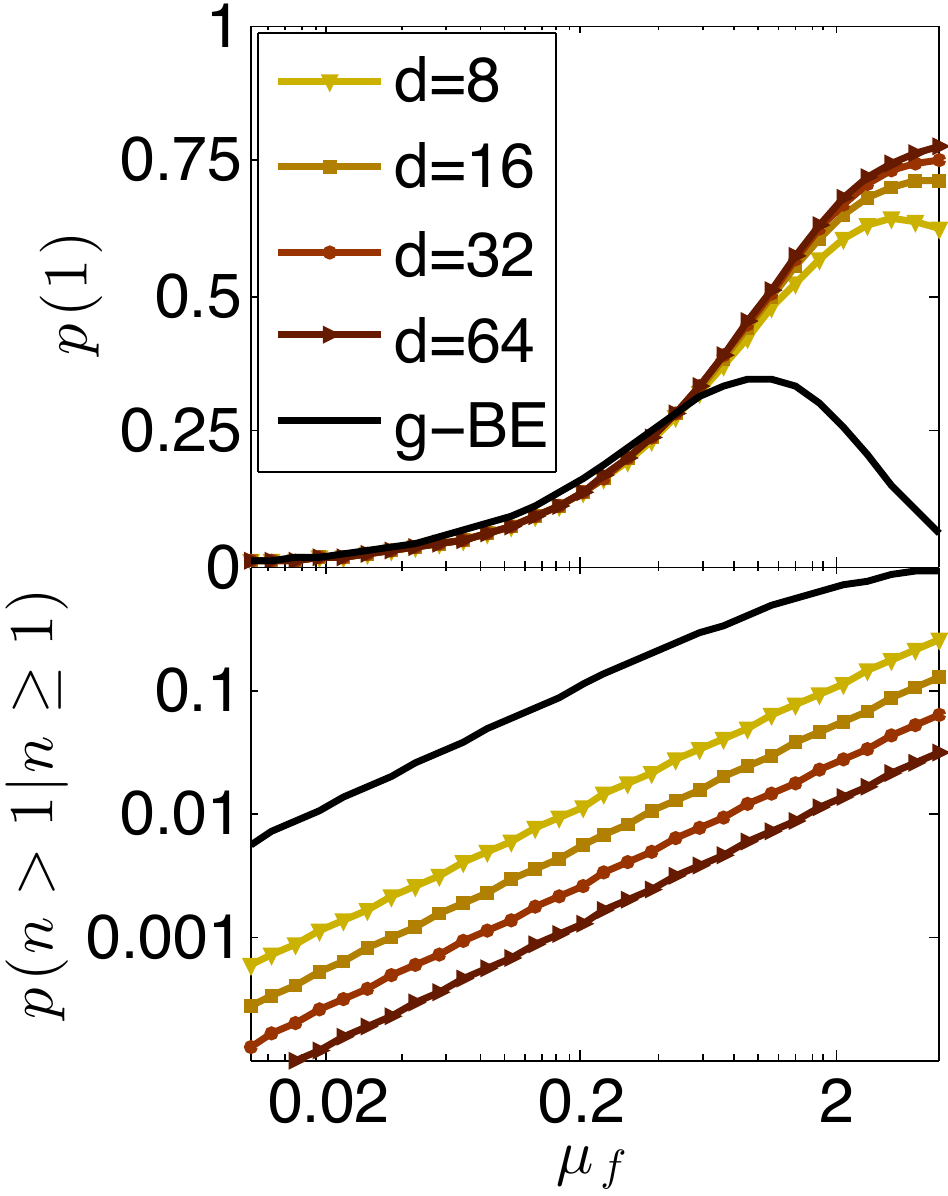}
\centering\caption{\small{(upper) The probability of Alice sending one photon per frame to Bob, $p(1)$, as a function of $\mu_f$ using the heralding circuit with the $d$ shown and for unheralded g-fold degenerate Bose-Einstein statistics, ``g-BE." (lower) The probability of Alice sending more than one photon per frame given that one or more photons were sent for the heralded and thermal statistics as above. We limit this probability to be less than $0.01$, which places an upper bound on $\mu_f$.}}
\label{herald}
\end{figure}

The protocol can only succeed if Alice registers a heralding event. She does so with probability $1-p_\text{fail}$, where
\begin{equation}
p_\text{fail}=\sum_{k=0}^\infty p(\mu_f,k,d)\cdot D_0(k).
\end{equation}
Here,
\begin{equation}
D_0(k)= (1-\eta_d)^k,
\end{equation}
where $\eta_d$ is the single photon detector efficiency, is the probability that $0$ out of $k$ photons are detected.

From this, the probability of generating $m$ photons on the output is 
\begin{equation}
\begin{split}
 p_H(m) = (1-p_\text{fail})&\sum_{k=m}^\infty p(\mu_b,k,d)D(k)\\
 &+ p_{\text{fail}}\delta_{m,0}
\end{split}
\label{eq:pH}
\end{equation}
where $\mu_b$ is the expected number of pairs generated in each time bin and $\delta_{x,y}$ is the Kronecker delta function.

In Eq. \ref{eq:pH},
\begin{equation}
D(k)=\binom{k}{m}\eta_s^m(1-\eta_s)^{k-m},
\end{equation}
where $\eta_s$ is the transmissivity of the modulator in the `on' position, is the probability that $m$ of the $k$ generated photons make it through the switching element.

We plot the results of this analysis in Fig. \ref{herald}. As $d$ increases for a given $\mu_f$, the average pair generation rate per time bin decreases resulting in a suppression of multiphoton emission during the frame. By limiting the probability of emitting multiple photons per frame given one or more photons were emitted, we determine the probability for generating one photon, $p(1)\approx p_\nu$. For $d=\{64,32,16,8\}$, $p_\nu =\{0.607,0.411,0.231,0.119\}$, respectively.

\bibliographystyle{apsrev_no_links.bst}


\begin{thebibliography}{37}
\expandafter\ifx\csname natexlab\endcsname\relax\def\natexlab#1{#1}\fi
\expandafter\ifx\csname bibnamefont\endcsname\relax
  \def\bibnamefont#1{#1}\fi
\expandafter\ifx\csname bibfnamefont\endcsname\relax
  \def\bibfnamefont#1{#1}\fi
\expandafter\ifx\csname citenamefont\endcsname\relax
  \def\citenamefont#1{#1}\fi
\expandafter\ifx\csname url\endcsname\relax
  \def\url#1{\texttt{#1}}\fi
\expandafter\ifx\csname urlprefix\endcsname\relax\def\urlprefix{URL }\fi
\providecommand{\bibinfo}[2]{#2}
\providecommand{\eprint}[2][]{\url{#2}}

\bibitem[{\citenamefont{Bennett and Brassard}(1984)}]{BB84}
\bibinfo{author}{\bibfnamefont{C.}~\bibnamefont{Bennett}} \bibnamefont{and}
  \bibinfo{author}{\bibfnamefont{G.}~\bibnamefont{Brassard}}, in
  \emph{\bibinfo{booktitle}{Proceedings of IEEE International Conference on
  Computers, Systems, and Signal Processing}} (\bibinfo{organization}{IEEE},
  \bibinfo{address}{New York}, \bibinfo{year}{1984}), pp.
  \bibinfo{pages}{175--179}.

\bibitem[{\citenamefont{Ekert}(1991)}]{1991.PRL.Ekert}
\bibinfo{author}{\bibfnamefont{A.~K.} \bibnamefont{Ekert}},
  \bibinfo{journal}{Phys. Rev. Lett.} \textbf{\bibinfo{volume}{67}},
  \bibinfo{pages}{661} (\bibinfo{year}{1991}).

\bibitem[{\citenamefont{Cerf et~al.}(2002)\citenamefont{Cerf, Bourennane,
  Karlsson, and Gisin}}]{2002.PRL.Cerf-Gisin.d-levelQKD}
\bibinfo{author}{\bibfnamefont{N.~J.} \bibnamefont{Cerf}},
  \bibinfo{author}{\bibfnamefont{M.}~\bibnamefont{Bourennane}},
  \bibinfo{author}{\bibfnamefont{A.}~\bibnamefont{Karlsson}}, \bibnamefont{and}
  \bibinfo{author}{\bibfnamefont{N.}~\bibnamefont{Gisin}},
  \bibinfo{journal}{Phys. Rev. Lett.} \textbf{\bibinfo{volume}{88}},
  \bibinfo{pages}{127902} (\bibinfo{year}{2002}).

\bibitem[{\citenamefont{Zhang et~al.}(2008)\citenamefont{Zhang, Silberhorn, and
  Walmsley}}]{PhysRevLett.100.110504}
\bibinfo{author}{\bibfnamefont{L.}~\bibnamefont{Zhang}},
  \bibinfo{author}{\bibfnamefont{C.}~\bibnamefont{Silberhorn}},
  \bibnamefont{and} \bibinfo{author}{\bibfnamefont{I.~A.}
  \bibnamefont{Walmsley}}, \bibinfo{journal}{Phys. Rev. Lett.}
  \textbf{\bibinfo{volume}{100}}, \bibinfo{pages}{110504}
  (\bibinfo{year}{2008}).

\bibitem[{\citenamefont{Ali~Khan and Howell}(2006)}]{PhysRevA.73.031801}
\bibinfo{author}{\bibfnamefont{I.}~\bibnamefont{Ali~Khan}} \bibnamefont{and}
  \bibinfo{author}{\bibfnamefont{J.~C.} \bibnamefont{Howell}},
  \bibinfo{journal}{Phys. Rev. A} \textbf{\bibinfo{volume}{73}},
  \bibinfo{pages}{031801} (\bibinfo{year}{2006}).

\bibitem[{\citenamefont{Bechmann-Pasquinucci and
  Tittel}(2000)}]{PhysRevA.61.062308}
\bibinfo{author}{\bibfnamefont{H.}~\bibnamefont{Bechmann-Pasquinucci}}
  \bibnamefont{and} \bibinfo{author}{\bibfnamefont{W.}~\bibnamefont{Tittel}},
  \bibinfo{journal}{Phys. Rev. A} \textbf{\bibinfo{volume}{61}},
  \bibinfo{pages}{062308} (\bibinfo{year}{2000}).

\bibitem[{\citenamefont{Ali-Khan et~al.}(2007)\citenamefont{Ali-Khan,
  Broadbent, and Howell}}]{PhysRevLett.98.060503}
\bibinfo{author}{\bibfnamefont{I.}~\bibnamefont{Ali-Khan}},
  \bibinfo{author}{\bibfnamefont{C.~J.} \bibnamefont{Broadbent}},
  \bibnamefont{and} \bibinfo{author}{\bibfnamefont{J.~C.}
  \bibnamefont{Howell}}, \bibinfo{journal}{Phys. Rev. Lett.}
  \textbf{\bibinfo{volume}{98}}, \bibinfo{pages}{060503}
  (\bibinfo{year}{2007}).

\bibitem[{\citenamefont{Tittel et~al.}(2000)\citenamefont{Tittel, Brendel,
  Zbinden, and Gisin}}]{PhysRevLett.84.4737}
\bibinfo{author}{\bibfnamefont{W.}~\bibnamefont{Tittel}},
  \bibinfo{author}{\bibfnamefont{J.}~\bibnamefont{Brendel}},
  \bibinfo{author}{\bibfnamefont{H.}~\bibnamefont{Zbinden}}, \bibnamefont{and}
  \bibinfo{author}{\bibfnamefont{N.}~\bibnamefont{Gisin}},
  \bibinfo{journal}{Phys. Rev. Lett.} \textbf{\bibinfo{volume}{84}},
  \bibinfo{pages}{4737} (\bibinfo{year}{2000}).

\bibitem[{\citenamefont{Qi}(2006)}]{Qi:06}
\bibinfo{author}{\bibfnamefont{B.}~\bibnamefont{Qi}}, \bibinfo{journal}{Opt.
  Lett.} \textbf{\bibinfo{volume}{31}}, \bibinfo{pages}{2795}
  (\bibinfo{year}{2006}).

\bibitem[{\citenamefont{Mower et~al.}()\citenamefont{Mower, Wong, Shapiro, and
  Englund}}]{DWDM_arxiv}
\bibinfo{author}{\bibfnamefont{J.}~\bibnamefont{Mower}},
  \bibinfo{author}{\bibfnamefont{F.}~\bibnamefont{Wong}},
  \bibinfo{author}{\bibfnamefont{J.}~\bibnamefont{Shapiro}}, \bibnamefont{and}
  \bibinfo{author}{\bibfnamefont{D.}~\bibnamefont{Englund}},
  \eprint{arxiv:1110.4867 [quant-ph]}.

\bibitem[{\citenamefont{Vaziri et~al.}(2002)\citenamefont{Vaziri, Weihs, and
  Zeilinger}}]{PhysRevLett.89.240401}
\bibinfo{author}{\bibfnamefont{A.}~\bibnamefont{Vaziri}},
  \bibinfo{author}{\bibfnamefont{G.}~\bibnamefont{Weihs}}, \bibnamefont{and}
  \bibinfo{author}{\bibfnamefont{A.}~\bibnamefont{Zeilinger}},
  \bibinfo{journal}{Phys. Rev. Lett.} \textbf{\bibinfo{volume}{89}},
  \bibinfo{pages}{240401} (\bibinfo{year}{2002}).

\bibitem[{\citenamefont{Mair et~al.}(2001)\citenamefont{Mair, Vaziri, Weihs,
  and Zeilinger}}]{Mair:2001fk}
\bibinfo{author}{\bibfnamefont{A.}~\bibnamefont{Mair}},
  \bibinfo{author}{\bibfnamefont{A.}~\bibnamefont{Vaziri}},
  \bibinfo{author}{\bibfnamefont{G.}~\bibnamefont{Weihs}}, \bibnamefont{and}
  \bibinfo{author}{\bibfnamefont{A.}~\bibnamefont{Zeilinger}},
  \bibinfo{journal}{Nature} \textbf{\bibinfo{volume}{412}},
  \bibinfo{pages}{313} (\bibinfo{year}{2001}).

\bibitem[{\citenamefont{Zhong et~al.}(2009)\citenamefont{Zhong, Wong, Roberts,
  and Battle}}]{Zhong:09}
\bibinfo{author}{\bibfnamefont{T.}~\bibnamefont{Zhong}},
  \bibinfo{author}{\bibfnamefont{F.~N.~C.} \bibnamefont{Wong}},
  \bibinfo{author}{\bibfnamefont{T.~D.} \bibnamefont{Roberts}},
  \bibnamefont{and} \bibinfo{author}{\bibfnamefont{P.}~\bibnamefont{Battle}},
  \bibinfo{journal}{Opt. Express} \textbf{\bibinfo{volume}{17}},
  \bibinfo{pages}{12019} (\bibinfo{year}{2009}).

\bibitem[{\citenamefont{Law and Eberly}(2004)}]{PhysRevLett.92.127903}
\bibinfo{author}{\bibfnamefont{C.~K.} \bibnamefont{Law}} \bibnamefont{and}
  \bibinfo{author}{\bibfnamefont{J.~H.} \bibnamefont{Eberly}},
  \bibinfo{journal}{Phys. Rev. Lett.} \textbf{\bibinfo{volume}{92}},
  \bibinfo{pages}{127903} (\bibinfo{year}{2004}).

\bibitem[{\citenamefont{Osgood et~al.}(2009)\citenamefont{Osgood, Panoiu,
  Dadap, Liu, Chen, Hsieh, Dulkeith, Green, and Vlasov}}]{Engin}
\bibinfo{author}{\bibfnamefont{R.~M.} \bibnamefont{Osgood}},
  \bibinfo{author}{\bibfnamefont{N.~C.} \bibnamefont{Panoiu}},
  \bibinfo{author}{\bibfnamefont{J.~I.} \bibnamefont{Dadap}},
  \bibinfo{author}{\bibfnamefont{X.}~\bibnamefont{Liu}},
  \bibinfo{author}{\bibfnamefont{X.}~\bibnamefont{Chen}},
  \bibinfo{author}{\bibfnamefont{I.-W.} \bibnamefont{Hsieh}},
  \bibinfo{author}{\bibfnamefont{E.}~\bibnamefont{Dulkeith}},
  \bibinfo{author}{\bibfnamefont{W.~M.} \bibnamefont{Green}}, \bibnamefont{and}
  \bibinfo{author}{\bibfnamefont{Y.~A.} \bibnamefont{Vlasov}},
  \bibinfo{journal}{Adv. Opt. Photon.} \textbf{\bibinfo{volume}{1}},
  \bibinfo{pages}{162} (\bibinfo{year}{2009}).

\bibitem[{\citenamefont{Franson}(1992)}]{PhysRevA.45.3126}
\bibinfo{author}{\bibfnamefont{J.~D.} \bibnamefont{Franson}},
  \bibinfo{journal}{Phys. Rev. A} \textbf{\bibinfo{volume}{45}},
  \bibinfo{pages}{3126} (\bibinfo{year}{1992}).

\bibitem[{\citenamefont{Assefa and Vlasov}(2007)}]{Assefa:07}
\bibinfo{author}{\bibfnamefont{S.}~\bibnamefont{Assefa}} \bibnamefont{and}
  \bibinfo{author}{\bibfnamefont{Y.~A.} \bibnamefont{Vlasov}},
  \bibinfo{journal}{Opt. Express} \textbf{\bibinfo{volume}{15}},
  \bibinfo{pages}{17562} (\bibinfo{year}{2007}).

\bibitem[{\citenamefont{Madsen et~al.}(1999)\citenamefont{Madsen, Lenz, Bruce,
  Cappuzzo, Gomez, and Scotti}}]{806867}
\bibinfo{author}{\bibfnamefont{C.}~\bibnamefont{Madsen}},
  \bibinfo{author}{\bibfnamefont{G.}~\bibnamefont{Lenz}},
  \bibinfo{author}{\bibfnamefont{A.}~\bibnamefont{Bruce}},
  \bibinfo{author}{\bibfnamefont{M.}~\bibnamefont{Cappuzzo}},
  \bibinfo{author}{\bibfnamefont{L.}~\bibnamefont{Gomez}}, \bibnamefont{and}
  \bibinfo{author}{\bibfnamefont{R.}~\bibnamefont{Scotti}},
  \bibinfo{journal}{Photonics Technology Letters, IEEE}
  \textbf{\bibinfo{volume}{11}}, \bibinfo{pages}{1623 } (\bibinfo{year}{1999}).

\bibitem[{\citenamefont{Deutsch et~al.}(1996)\citenamefont{Deutsch, Ekert,
  Jozsa, Macchiavello, Popescu, and Sanpera}}]{PhysRevLett.77.2818}
\bibinfo{author}{\bibfnamefont{D.}~\bibnamefont{Deutsch}},
  \bibinfo{author}{\bibfnamefont{A.}~\bibnamefont{Ekert}},
  \bibinfo{author}{\bibfnamefont{R.}~\bibnamefont{Jozsa}},
  \bibinfo{author}{\bibfnamefont{C.}~\bibnamefont{Macchiavello}},
  \bibinfo{author}{\bibfnamefont{S.}~\bibnamefont{Popescu}}, \bibnamefont{and}
  \bibinfo{author}{\bibfnamefont{A.}~\bibnamefont{Sanpera}},
  \bibinfo{journal}{Phys. Rev. Lett.} \textbf{\bibinfo{volume}{77}},
  \bibinfo{pages}{2818} (\bibinfo{year}{1996}).

\bibitem[{\citenamefont{Devetak and Winter}(2005)}]{Devetak08012005}
\bibinfo{author}{\bibfnamefont{I.}~\bibnamefont{Devetak}} \bibnamefont{and}
  \bibinfo{author}{\bibfnamefont{A.}~\bibnamefont{Winter}},
  \bibinfo{journal}{Proceedings of the Royal Society A: Mathematical, Physical
  and Engineering Science} \textbf{\bibinfo{volume}{461}}, \bibinfo{pages}{207}
  (\bibinfo{year}{2005}).

\bibitem[{\citenamefont{Garc\'ia-Patr\'on and
  Cerf}(2006)}]{PhysRevLett.97.190503}
\bibinfo{author}{\bibfnamefont{R.}~\bibnamefont{Garc\'ia-Patr\'on}}
  \bibnamefont{and} \bibinfo{author}{\bibfnamefont{N.~J.} \bibnamefont{Cerf}},
  \bibinfo{journal}{Phys. Rev. Lett.} \textbf{\bibinfo{volume}{97}},
  \bibinfo{pages}{190503} (\bibinfo{year}{2006}).

\bibitem[{\citenamefont{Holevo}(1973)}]{holevo73}
\bibinfo{author}{\bibfnamefont{A.}~\bibnamefont{Holevo}},
  \bibinfo{journal}{Probl. Peredachi Inf.} \textbf{\bibinfo{volume}{9}},
  \bibinfo{pages}{3} (\bibinfo{year}{1973}).

\bibitem[{\citenamefont{Navascu\'es et~al.}(2006)\citenamefont{Navascu\'es,
  Grosshans, and Ac\'in}}]{PhysRevLett.97.190502}
\bibinfo{author}{\bibfnamefont{M.}~\bibnamefont{Navascu\'es}},
  \bibinfo{author}{\bibfnamefont{F.}~\bibnamefont{Grosshans}},
  \bibnamefont{and} \bibinfo{author}{\bibfnamefont{A.}~\bibnamefont{Ac\'in}},
  \bibinfo{journal}{Phys. Rev. Lett.} \textbf{\bibinfo{volume}{97}},
  \bibinfo{pages}{190502} (\bibinfo{year}{2006}).

\bibitem[{\citenamefont{Holevo et~al.}(1999)\citenamefont{Holevo, Sohma, and
  Hirota}}]{PhysRevA.59.1820}
\bibinfo{author}{\bibfnamefont{A.~S.} \bibnamefont{Holevo}},
  \bibinfo{author}{\bibfnamefont{M.}~\bibnamefont{Sohma}}, \bibnamefont{and}
  \bibinfo{author}{\bibfnamefont{O.}~\bibnamefont{Hirota}},
  \bibinfo{journal}{Phys. Rev. A} \textbf{\bibinfo{volume}{59}},
  \bibinfo{pages}{1820} (\bibinfo{year}{1999}).

\bibitem[{\citenamefont{Serafini et~al.}(2004)\citenamefont{Serafini,
  Illuminati, and Siena}}]{0953-4075-37-2-L02}
\bibinfo{author}{\bibfnamefont{A.}~\bibnamefont{Serafini}},
  \bibinfo{author}{\bibfnamefont{F.}~\bibnamefont{Illuminati}},
  \bibnamefont{and} \bibinfo{author}{\bibfnamefont{S.~D.} \bibnamefont{Siena}},
  \bibinfo{journal}{Journal of Physics B: Atomic, Molecular and Optical
  Physics} \textbf{\bibinfo{volume}{37}}, \bibinfo{pages}{L21}
  (\bibinfo{year}{2004}).

\bibitem[{\citenamefont{Weedbrook et~al.}(2012)\citenamefont{Weedbrook,
  Pirandola, Garc\'ia-Patr\'on, Cerf, Ralph, Shapiro, and
  Lloyd}}]{RevModPhys.84.621}
\bibinfo{author}{\bibfnamefont{C.}~\bibnamefont{Weedbrook}},
  \bibinfo{author}{\bibfnamefont{S.}~\bibnamefont{Pirandola}},
  \bibinfo{author}{\bibfnamefont{R.}~\bibnamefont{Garc\'ia-Patr\'on}},
  \bibinfo{author}{\bibfnamefont{N.~J.} \bibnamefont{Cerf}},
  \bibinfo{author}{\bibfnamefont{T.~C.} \bibnamefont{Ralph}},
  \bibinfo{author}{\bibfnamefont{J.~H.} \bibnamefont{Shapiro}},
  \bibnamefont{and} \bibinfo{author}{\bibfnamefont{S.}~\bibnamefont{Lloyd}},
  \bibinfo{journal}{Rev. Mod. Phys.} \textbf{\bibinfo{volume}{84}},
  \bibinfo{pages}{621} (\bibinfo{year}{2012}).

\bibitem[{\citenamefont{Delgado and Muga}(1997)}]{PhysRevA.56.3425}
\bibinfo{author}{\bibfnamefont{V.}~\bibnamefont{Delgado}} \bibnamefont{and}
  \bibinfo{author}{\bibfnamefont{J.~G.} \bibnamefont{Muga}},
  \bibinfo{journal}{Phys. Rev. A} \textbf{\bibinfo{volume}{56}},
  \bibinfo{pages}{3425} (\bibinfo{year}{1997}).

\bibitem[{\citenamefont{Mandel and Wolf}(1995)}]{MandelandWolf}
\bibinfo{author}{\bibfnamefont{L.}~\bibnamefont{Mandel}} \bibnamefont{and}
  \bibinfo{author}{\bibfnamefont{E.}~\bibnamefont{Wolf}},
  \emph{\bibinfo{title}{Optical Coherence and Quantum Optics}},
  \bibinfo{number}{Section 12.11} (\bibinfo{publisher}{Cambridge University
  Press}, \bibinfo{year}{1995}), \bibinfo{edition}{1st} ed.

\bibitem[{\citenamefont{Jouguet et~al.}(2013)\citenamefont{Jouguet,
  Kunz-Jacques, Leverrier, Grangier, and Diamanti}}]{80km_cvqkd}
\bibinfo{author}{\bibfnamefont{P.}~\bibnamefont{Jouguet}},
  \bibinfo{author}{\bibfnamefont{S.}~\bibnamefont{Kunz-Jacques}},
  \bibinfo{author}{\bibfnamefont{A.}~\bibnamefont{Leverrier}},
  \bibinfo{author}{\bibfnamefont{P.}~\bibnamefont{Grangier}}, \bibnamefont{and}
  \bibinfo{author}{\bibfnamefont{E.}~\bibnamefont{Diamanti}},
  \bibinfo{journal}{Nature Photonics} \textbf{\bibinfo{volume}{7}},
  \bibinfo{pages}{378} (\bibinfo{year}{2013}).

\bibitem[{\citenamefont{Pittman et~al.}(2002)\citenamefont{Pittman, Jacobs, and
  Franson}}]{PhysRevA.66.042303}
\bibinfo{author}{\bibfnamefont{T.~B.} \bibnamefont{Pittman}},
  \bibinfo{author}{\bibfnamefont{B.~C.} \bibnamefont{Jacobs}},
  \bibnamefont{and} \bibinfo{author}{\bibfnamefont{J.~D.}
  \bibnamefont{Franson}}, \bibinfo{journal}{Phys. Rev. A}
  \textbf{\bibinfo{volume}{66}}, \bibinfo{pages}{042303}
  (\bibinfo{year}{2002}).

\bibitem[{\citenamefont{Mower and Englund}(2011)}]{PhysRevA.84.052326}
\bibinfo{author}{\bibfnamefont{J.}~\bibnamefont{Mower}} \bibnamefont{and}
  \bibinfo{author}{\bibfnamefont{D.}~\bibnamefont{Englund}},
  \bibinfo{journal}{Phys. Rev. A} \textbf{\bibinfo{volume}{84}},
  \bibinfo{pages}{052326} (\bibinfo{year}{2011}).

\bibitem[{\citenamefont{Lo et~al.}(2005)\citenamefont{Lo, Ma, and
  Chen}}]{PhysRevLett.94.230504}
\bibinfo{author}{\bibfnamefont{H.-K.} \bibnamefont{Lo}},
  \bibinfo{author}{\bibfnamefont{X.}~\bibnamefont{Ma}}, \bibnamefont{and}
  \bibinfo{author}{\bibfnamefont{K.}~\bibnamefont{Chen}},
  \bibinfo{journal}{Phys. Rev. Lett.} \textbf{\bibinfo{volume}{94}},
  \bibinfo{pages}{230504} (\bibinfo{year}{2005}).

\bibitem[{\citenamefont{Renner et~al.}(2005)\citenamefont{Renner, Gisin, and
  Kraus}}]{PhysRevA.72.012332}
\bibinfo{author}{\bibfnamefont{R.}~\bibnamefont{Renner}},
  \bibinfo{author}{\bibfnamefont{N.}~\bibnamefont{Gisin}}, \bibnamefont{and}
  \bibinfo{author}{\bibfnamefont{B.}~\bibnamefont{Kraus}},
  \bibinfo{journal}{Phys. Rev. A} \textbf{\bibinfo{volume}{72}},
  \bibinfo{pages}{012332} (\bibinfo{year}{2005}).

\bibitem[{\citenamefont{Mandel}(1959)}]{Mandelphotstat}
\bibinfo{author}{\bibfnamefont{L.}~\bibnamefont{Mandel}}, in
  \emph{\bibinfo{booktitle}{Proceedings of the Physical Society}}
  (\bibinfo{year}{1959}), vol.~\bibinfo{volume}{74}, pp.
  \bibinfo{pages}{233--24}.

\bibitem[{\citenamefont{Yang et~al.}(2011)\citenamefont{Yang, Ma, Guo, Cui, and
  Li}}]{Yang:2011fk}
\bibinfo{author}{\bibfnamefont{L.}~\bibnamefont{Yang}},
  \bibinfo{author}{\bibfnamefont{X.}~\bibnamefont{Ma}},
  \bibinfo{author}{\bibfnamefont{X.}~\bibnamefont{Guo}},
  \bibinfo{author}{\bibfnamefont{L.}~\bibnamefont{Cui}}, \bibnamefont{and}
  \bibinfo{author}{\bibfnamefont{X.}~\bibnamefont{Li}} (\bibinfo{year}{2011}).

\bibitem[{\citenamefont{Marsili et~al.}(2013)\citenamefont{Marsili, Verma,
  Stern, Harrington, Lita, Gerrits, Vayshenker, Baek, Shaw, Mirin
  et~al.}}]{sw-nam93}
\bibinfo{author}{\bibfnamefont{F.}~\bibnamefont{Marsili}},
  \bibinfo{author}{\bibfnamefont{B.}~\bibnamefont{Verma}},
  \bibinfo{author}{\bibfnamefont{A.}~\bibnamefont{Stern}},
  \bibinfo{author}{\bibfnamefont{S.}~\bibnamefont{Harrington}},
  \bibinfo{author}{\bibfnamefont{A.}~\bibnamefont{Lita}},
  \bibinfo{author}{\bibfnamefont{T.}~\bibnamefont{Gerrits}},
  \bibinfo{author}{\bibfnamefont{I.}~\bibnamefont{Vayshenker}},
  \bibinfo{author}{\bibfnamefont{B.}~\bibnamefont{Baek}},
  \bibinfo{author}{\bibfnamefont{M.}~\bibnamefont{Shaw}},
  \bibinfo{author}{\bibfnamefont{R.}~\bibnamefont{Mirin}},
  \bibnamefont{et~al.}, \bibinfo{journal}{Nature Photonics}
  \textbf{\bibinfo{volume}{7}}, \bibinfo{pages}{210} (\bibinfo{year}{2013}).

\bibitem[{\citenamefont{Dauler et~al.}(2007)\citenamefont{Dauler, Robinson,
  Kerman, Yang, Rosfjord, Anant, Voronov, Gol'tsman, and Berggren}}]{4277352}
\bibinfo{author}{\bibfnamefont{E.}~\bibnamefont{Dauler}},
  \bibinfo{author}{\bibfnamefont{B.}~\bibnamefont{Robinson}},
  \bibinfo{author}{\bibfnamefont{A.}~\bibnamefont{Kerman}},
  \bibinfo{author}{\bibfnamefont{J.~K.~W.} \bibnamefont{Yang}},
  \bibinfo{author}{\bibfnamefont{K.}~\bibnamefont{Rosfjord}},
  \bibinfo{author}{\bibfnamefont{V.}~\bibnamefont{Anant}},
  \bibinfo{author}{\bibfnamefont{B.}~\bibnamefont{Voronov}},
  \bibinfo{author}{\bibfnamefont{G.}~\bibnamefont{Gol'tsman}},
  \bibnamefont{and} \bibinfo{author}{\bibfnamefont{K.}~\bibnamefont{Berggren}},
  \bibinfo{journal}{Applied Superconductivity, IEEE Transactions on}
  \textbf{\bibinfo{volume}{17}}, \bibinfo{pages}{279} (\bibinfo{year}{2007}).

\end{thebibliography}
\end{document}